%% file: copenhagentalk.tex
\documentclass[12pt]{article}
\usepackage{epsfig}

\input econfmacros.tex
\def\Title#1{\begin{center} {\Large {\bf #1} } \end{center}}

\begin{document}

\Title{Introduction to Effective Lagrangians for QCD}

\bigskip\bigskip


\begin{raggedright}  

{\it Joseph Schechter\index{Schechter, J.}\\
Department of Physics\\
Syracuse University\\
Syracuse, New York 13244-1130, USA}
\end{raggedright}

\begin{abstract}
    A brief introduction to the effective Lagrangian
treatment of QCD (in the
sense of using fields representing physical particles rather than quarks
and gluons) will be given. The historical evolution of the subject
will be discussed. Some background material related to a recent model for
Gamma Ray Bursters will be given. Finally, some recent work on low energy 
strong interactions will be mentioned.
\end{abstract}

 \section{Need for an effective Lagrangian; historical background}
     There is little doubt that, at least up to energies probed by
present accelerators, QCD is the correct theory of strong interactions.
It contains the three light spin 1/2 quark fields
  u, d and s as well
as the three heavy quark fields c, b and t. We will focus attention
completely on the light quarks here. The dynamics is governed by
an SU(3) Yang Mills~\cite{ym} theory; roughly this means that the strong
force 
has a general similarity to the ordinary QED force which involves the
exchange
of a gauge field (photon). However there are eight ``photons" rather than
a single one in QCD. They may be put into a traceless $3 \times 3$
matrix, $A_\mu$ and the interaction term in the fundamental Lagrangian
for, say, the $u$ quark is $ig{\bar u}\gamma_{\mu}A_{\mu}u$, where $g$
is a coupling constant. Notice that $u$ is a column vector with unwritten
"color" indices. Unlike the QED case, the  QCD ``photons" have  self-
interaction terms with the structures
$-igTr({\partial}_{\mu}A_\nu [A_\mu,A_\nu])+
\frac{g^2}{4}Tr([A_\mu,A_\nu][A_\mu,A_\nu])$. These have the consequence
at the first loop order of perturbation theory that the effective
energy dependent coupling constant behaves as
\begin{equation}
\frac{g^2(E)}{4\pi} \sim 1/ln(\frac{E}{\Lambda}) ,
\end{equation} 
where $E$ is the energy scale at which the theory is being applied and
$\Lambda$ is a fixed number of order 250 MeV characterizing QCD. At high
energies (above several GeV) $g$ is small and perturbation theory is
good. However at low energies, where we want to discuss topics like 
binding of quarks to make mesons, interactions of light mesons, CP
violation in  K meson decays, Nuclear Physics, etc. etc., g is large so
QCD is non perturbative. Then one must adopt some other approach.
It seems reasonable to hope that this strong coupling at low energies
tightly binds the quarks into mesons which interact weakly enough
among themselves to be described by the perturbative treatment of an 
{\it effective Lagrangian} constructed from the light meson fields
\footnote{Baryons can either be included directly or observed to emerge  
as solitons\cite{sk} in the effective meson theory.} rather than the
quarks and gluons. This hope turns out to be realized.

    The crucial idea in constructing the effective Lagrangian
is to mock up the symmetries observed in nature (and which are displayed 
by the more fundamental QCD Lagrangian). Apart from the Lorentz and
discrete space-time symmetries of the strong interaction, the starting point
is the imposition of Wigner's isotopic spin symmetry, now denoted $SU(2)_V$. The proton
and
neutron belong to a spinor of $SU(2)_V$ while three pion
fields representing linear combinations of the three pion charge states
belong to a vector. The corresponding transformations are:
\begin{equation}
SU(2)_V :\quad  N=\pmatrix{p\cr n} \rightarrow U_VN \quad
{\rm and} \quad \phi=\frac{1}{\sqrt{2}}\mbox{\boldmath
 $\pi\cdot\tau$} \rightarrow
U_{V}\phi{U_V}^\dagger,
\label{ispin}
\end{equation}
      where $U_V$ is a two dimensional unitary unimodular matrix. The subscript
$V$ stands for vector. Yukawa's original theory implies
 an effective Lagrangian with the $SU(2)_V$ invariant interaction terms:
\begin{equation}
ig_Y{\bar N} \phi\gamma_5 N + \lambda[Tr(\phi^2)]^2 .
\label{yukawa}
\end{equation}  
Here $g_Y$ is the Yukawa coupling constant while $\lambda$ is the coupling
constant
for a term which gives $\pi \pi $ scattering.

      Does the Yukawa Lagrangian actually work?
If it is to be a reasonable effective Lagrangian it should, like QED, give
reasonable results already at tree level. Now the value of $g_Y$ has been known
for about 60 years (determined from the long distance part of the nucleon nucleon
force due to pion exchange). Around 50 years ago pions were made in accelerators and
pion nucleon scattering measured. At tree level, the amplitude for this scattering
process should correspond to nucleon ``exchange" and be proportional to $g_Y^2$.
Unfortunately it turns out, near the scattering threshold, to be more than
an
order of magnitude larger than experiment. This would appear to doom
the Yukawa theory as 
the basis for a perturbative treatment. However, there is a small
modification
which saves it.
 
     Before getting to this modification we note that the discovery of strange
 particles
indicated that the isotopic symmetry group $SU(2)_V$ is too small and that
it should be upgraded to $SU(3)_V$ while the meson matrix $\phi$ in
eq.(\ref{ispin}) should 
be upgraded~\cite{ioo} to a $3 \times 3$ matrix. The mesons are now
pictured as
composites, $q{\bar q}$ of a quark and anti-quark~\cite{g,z}. Similarly
the baryons are
regarded as composites $qqq$. At the fundamental quark level the $SU(3)_V$ symmetry is
realized as:
\begin{equation}
q=\pmatrix{u\cr d\cr s} \rightarrow U_Vq ,
\label{su3}
\end{equation}
where $U_V$ is now a $3 \times 3$ unitary matrix. It was understood from the
beginning~\cite{g,ok} 
that the symmetry group $SU(3)_V$ has to be an approximate one; it could only be exact in a
world in which the nucleons and strange baryons like the $\Lambda$ are degenerate.  

    The desired modification of the Yukawa theory which saves the
perturbative pion nucleon scattering prediction arises from the extension of the isotopic
spin symmetry in another direction. This may be motivated by
consideration of the ``V-A" theory of beta decay type interactions, discovered \cite{vminusa}
a little more than forty years ago. In that well confirmed model, which displays maximal
parity violation, the fundamental
fermion fields (then nucleons but now quarks) enter into the interaction only through
their projection:
$q_L=\frac{1+\gamma_5}{2}q$ .
For a massless fermion, this corresponds to a left handed helicity state (momentum vector
opposite to spin vector). The strong interaction, which conserves
 parity, of course, requires
the inclusion also of the right handed projection
 $q_R = \frac{1-\gamma_5}{2}q$. What this
 seems to be telling us is that Nature likes to employ
 the projected (chiral) fields. Around 
that time, well before QCD was known, it raised the question as to whether Nature
also chooses to realize
 isotopic type symmetries on the chiral components. The chiral
(for generality) $U(3)_L \times U(3)_R$ symmetry corresponds to demanding invariance under
the transformations:
\begin{equation}
q_{L,R} \rightarrow U_{L,R}q_{L,R} ,
\label{chiraltransf}
\end{equation}
where $U_L$ and $U_R$ are two separate $3 \times 3 $ unitary matrices. This
 symmetry
is an approximate one since it is manifestly broken by non zero fundamental
 fermion masses.
The vector symmetry corresponds to the special choice $U_L=U_R=U_V$.
 In order to implement
the chiral symmetry in the usual manner it is necessary 
to further enlarge the
 meson multiplet. In the simplest
version which uses $SU(2)_L \times SU(2)_R$ it means adding
 to the three pion field
components an isosinglet scalar field called the sigma~\cite{gl},
 whose coupling to the
nucleon is correlated with that of the pion. 
We will discuss this more soon but for the moment just note that
 it modifies the pion nucleon
scattering diagrams to those shown in Fig 1. The new sigma exchange
 diagram almost completely
cancels the contribution of the nucleon exchange diagrams. The much
 smaller answer  agrees
 at low energies with
experiment to about 15 per cent, if the mass of the sigma is considerably heavier
than that of the pions. Other similarly good strong interaction predictions
can be made by the same model. Thus it seems that chiral symmetry has
``saved" the Yukawa theory as the effective description of strong
interactions at low energies.

\begin{figure}[htb]
\begin{center}
\epsfig{file=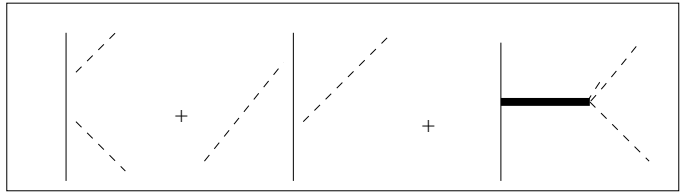,height=1in}
\caption{Nucleon=full line,pion=dashed line, sigma=heavy line.}
\label{fig:pin}
\end{center}
\end{figure}

\section{Effective Lagrangian of mesons}

One of the many questions raised by the remarks above 
concerns the consistency of the
requirement
that the mass of the sigma be appreciably greater than the mass of the 
pion,  
with the requirement that they both belong to the same chiral multiplet
and hence should be, at least approximately, degenerate in mass. To investigate this
further it is easier to focus just on the mesons. The same problem of too large
predicted low energy scattering is displayed by pi pi scattering and the solution
of including sigma exchanges is also the same.

The meson field multiplet, $M_{ab}$ with the correct chiral properties is
 schematically constructed from
the underlying quark fields as:
\begin{equation}
{\bar q}_{Rb}q_{La} \sim M_{ab} = S_{ab} + i {\phi}_{ab} .
\label{Mfield}
\end{equation}
where the decomposition of $M$ into pseudoscalar, $\phi = {\phi}^\dagger$
and scalar, $S = S^\dagger$ pieces is shown. M has the transformation properties
\begin{equation}
U(3)_L \times U(3)_R : \quad M \rightarrow U_LM{U_R}^\dagger , \quad 
{\rm parity}: \quad M({\bf x}) \rightarrow M^{\dagger}(-{\bf x}).
\label{Mtransf}
\end{equation}
One may check these by noting that the Yukawa like term $ {\bar q}_L M q_R + h.c.$
is invariant. The simplest Lagrangian made from $M$ is:
\begin{equation}
{\cal L}_{meson} = -\frac{1}{2}Tr({\partial}_\mu M {\partial}_\mu
M^{\dagger})
-V_0(M,M^{\dagger}) +{\sum}A_a(M_{aa}+{M^\dagger}_{aa}).
\label{lmeson}
\end{equation}
The first term is the standard kinetic term while the second term
might as well be taken to be the most general non derivative function of the
independent 
chiral $SU(3)_L \times SU(3)_R$ invariants $I_1=Tr(MM^\dagger)$, $I_2=Tr[(MM^\dagger)^2]$,
$I_3=Tr[(MM^\dagger)^3]$ and $I_4= 6(detM+detM^\dagger)$. The last term provides the
required chiral as well as flavor (vector) type symmetry breaking, with the real
numbers $A_a$ being proportional to the masses $m_a$ of the light quarks. It turns
out~\cite{su1} that many of the most interesting consequences
 of eq.(\ref{lmeson})
at tree level
are independent of the specific form of $V_0$; only its symmetry
 properties are needed. The consequences of chiral symmetry
are relations between $n$ and $n-1$ point vertices; for example,
trilinear vertices are related to masses while
masses are related to ``decay constants".  
Another point of interest is that the presence of the quantity $I_4$ in $V_0$
spoils the invariance under ``axial baryon number"
or $U(1)_A$
transformations of the form $M \rightarrow phase \times M$. If this extra symmetry were
allowed to
remain it would force~\cite{su2} one of the eta type mesons to be degenerate with the pi, in
clear contradiction to nature.

The names of the 18 particles belonging to $M$, which comprises a basis for the
irreducible representation $(3,3^*) + (3^*,3)$ of
 $SU(3)_L \times SU(3)_R$ and parity, are
listed in Table~\ref{Mstates}.

\begin{table}[b]
\begin{center}
\begin{tabular}{l|cccc}
$spin^{parity}$&$I=1$&$I=0$&$I=1/2$&$I=0$\\ \hline
$0^-$&$\mbox{\boldmath $\pi$}$&$\eta$&$K's$&${\eta}^\prime$\\
$0^+$&${\bf a}_0$&$\sigma$&${\kappa}'s$&${\sigma}^{\prime}$\\ \hline
\end{tabular}
\caption{States represented by $M$.}
\label{Mstates}
\end{center}
\end{table}

 It is helpful to first consider the two flavor
case wherein M is just a $2 \times 2$ matrix. This corresponds to keeping the 
eight particles in the first two columns of Table~\ref{Mstates}. However a further
simplification is possible since, for only $N=2$,
 the fundamental representation of $SU(N)$ is equivalent to its complex conjugate.
Using the fact that $U{\tau}_2={\tau}_2U^*$ for any 2 dimensional
 unitary, unimodular matrix, $U$ we observe that under an $SU(2)_L\times SU(2)_R$
transformation, ${\tau}_2M^*{\tau}_2 \rightarrow U_L({\tau}_2M^*{\tau}_2){U_R}^\dagger$.
This is the same transformation as in eq.(\ref{Mtransf}) so we have the following two
linear combinations which each transform irreducibly under $SU(2)_L \times SU(2)_R$:
\begin{equation}
\frac{1}{\sqrt 2}(M+{\tau}_2M^*{\tau}_2)=\sigma+i\mbox{\boldmath 
$\pi\cdot\tau$},
\quad
\frac{1}{\sqrt 2}(M-{\tau}_2M^*{\tau}_2)=i\eta+{\bf a}_0
\mbox{\boldmath $\cdot\tau$}.
\label{su2M}
\end{equation}
Thus it is consistent to construct a theory using just the $\sigma$ and $\pi$ fields;
this is the choice made in the Gell-Mann Levy model~\cite{gl} discussed 
above. 

The chiral invariant potential in the Gell-Mann Levy model
 is taken to be:
\begin{equation}
V_0=-b({\sigma}^2 + {\mbox{\boldmath $\pi$}}^2) + 
\lambda({\sigma}^2+{\mbox{\boldmath
$\pi$}}^2)^2,
\label{glpot}
\end{equation}
where $b$ and $\lambda$ are real, positive constants.
 Because of the ``wrong sign"
quadratic term, the parity conserving minimum of the potential
 will occur for
$\mbox{\boldmath $\pi$} = 0$ and $(\sigma)_{min} \ne 0$.
Once the Lagrangian is rewritten in terms of new "small
oscillation" fields defined as deviations from
their values at the potential minimum, it is evident that the resulting
theory no longer
 has the full chiral symmetry.
This is the familiar phenomenon of spontaneous symmetry breakdown.
 The remaining
symmetry is just isospin invariance ($SU(2)_V$) and, neglecting the quark
masses for simplicity,
the pion is forced to become a Nambu Goldstone
 boson (i.e. to have zero mass).
 On the other hand the mass of the sigma is a free parameter.
 This solves the problem of degenerate pion and sigma, mentioned
at the beginning of this section. 

At the level of fundamental (idealized to be massless) fermions,
it was pointed out~\cite{njl} that the origin of $(\sigma)_{min}
\ne 0$ is, analogously to the theory of superconductivity, a pairing
force between (in modern language) $q_L$ and ${\bar q}_R$. This leads
to a ground state $|0>$ which is a ``condensate" of
 such pairs and their conjugates, characterized by $<0|{\bar q}_aq_a|0>
\ne0$.

It is interesting to discuss the pi pi scattering amplitude in
this model. There is a $\sigma\phi\phi$ interaction as well as the
four pion self interaction term shown in eq.(\ref{yukawa}). The
amplitude for any particular choice of pion charges is 
conventionally specified by a suitable linear combination of the single
function $A(s,t,u)$
and its permutations under the interchange of the Mandelstam variables
$s$, $t$ and $u$. At tree level one has:
\begin{equation}
A(s,t,u) =\frac{2}{F_{\pi}^2}(m_{\sigma}^2-m_{\pi}^2)(\frac{m_{\sigma}^2
-m_{\pi}^2}{m_{\sigma}^2-s}-1) ,
\label{pipiamp}
\end{equation}
where the first term represents the sigma exchange piece and the
second term, the contact interaction piece. Furthermore the quantity
$F_{\pi}$ ($= {\sqrt 2}(\sigma)_{min}$ in the model) is identifiable
as the hadron factor of the amplitude for $\pi^- \rightarrow 
\mu^-{\bar \nu}_{\mu}$ and is numerically, $0.131$ GeV. At low
energies (i.e. just above $s=4m_{\pi}^2$ ) and considering $m_\sigma$
to be considerably larger than $m_\pi$, eq.(\ref{pipiamp}) becomes,
\begin{equation}
A(s,t,u) =\frac{2}{F_{\pi}^2}(s-m_{\pi}^2) ,
\label{caamp}
\end{equation}
 a formula due to Weinberg~\cite{w}. It is in reasonable agreement
with experiment and can be noted to emerge from the
difference
of the two terms in eq.(\ref{pipiamp}) which are each about twenty
times larger than the final result.

     While this numerical result is very encouraging, from the standpoint
of making a perturbation expansion it is not nice that it arises
from the near cancellation of two large quantities. Also, historically
the sigma meson has been hard to identify from experiment. For the latter
reason it was proposed already in~\cite{gl} to ``integrate out" the sigma
by imagining its mass to go to infinity. This may be done from the
equation of motion $\partial_\mu\partial_\mu \sigma=\frac{\partial
V_0}{\partial \sigma}$.
For an infinitely heavy particle we neglect the kinetic term so the 
equation of motion just becomes $0=\frac{\partial V_0}{\partial \sigma}$,
neglecting, at first, the ``quark mass terms".
Then eq.(\ref{glpot}) can be solved for $\sigma$:
\begin{equation}
\sigma ={\sqrt {F_{\pi}^2/2-\mbox{\boldmath $\pi$}^2}}.
\label{constraint}
\end{equation}
Substituting back into the two flavor version of eq.(\ref{lmeson}) yields
the
Gell-Mann Levy non linear sigma model:
\begin{equation}
{\cal L}=-\frac{1}{2}(\partial_\mu\mbox{\boldmath{$\pi$}})^2 
-\frac{1}{2}({\partial}_\mu \sqrt{F_{\pi}^2/2-\mbox{\boldmath 
{$\pi$}}^2}\\)^2. 
\label{glnlsm}
\end{equation}
Notice that the substitution of eq.(\ref{constraint}) replaces the
potential of eq.(\ref{glpot}) by a constant. All the interactions
in eq.(\ref{glnlsm}) involve only pions and are of derivative type.
This reproduces the desired result eq.(\ref{caamp}) directly without any
need to mention the sigma.

     It is possible to give a slightly more convenient form for this
model,
which was obtained independently~\cite{gn} of ref.~\cite{gl}. In terms of
the
unitary 
matrix ``chiral field" $U=exp(2i\phi/F_\pi)$ just write:
\begin{equation}
{\cal L}=
-\frac{F_{\pi}^2}{8}Tr(\partial_{\mu}U\partial_{\mu}U^{\dagger}).
\label{2fnlsm}
\end{equation}
Of course, the matrix, $U$ is defined by its power series expansion.
In order to verify the equivalence of eqs.(\ref{glnlsm}) and
 (\ref{2fnlsm}) one may make use of Chisholm's theorem~\cite{ct}. This
states that the transformation (analog of a point transformation in
classical mechanics):
\begin{equation}
\chi_a=\phi_a + \gamma_{abc}\phi_b\phi_c +\gamma_{abcd}\phi_b\phi_c\phi_d
+\cdots
\label{ct}
\end{equation}
between the sets of scalar fields $\phi_a$ and $\chi_a$ leads to
equivalent results at tree level.
An advantage of this formulation is that it can straightforwardly
be extended to the three flavor case just by considering $\phi$
in eq.(\ref{2fnlsm}) to be the $3\times3$ matrix of pseudoscalar fields.
Notice that $U$ transforms in the same (linear) way $M$ does
 in eq.(\ref{Mtransf}); this forces $\phi$ to transform non-linearly,
which give the model its name.
     
    A quick mnemonic for going from the three flavor linear model in
eq.(\ref{lmeson}) to the three flavor non-linear model is to first make the
``polar decomposition", $M=BU$ with $B$ hermitian and $U$ unitary,
and then replace $B$ by the the assumed $SU(3)$ symmetric ``vacuum" value
$<B>=\frac{F_\pi}{2}diag(1,1,1)$. Substituting this form into
eq.(\ref{lmeson})
yields finally:
\begin{equation}
 {\cal L}=-\frac{F_{\pi}^2}{8}Tr(\partial_{\mu}U\partial_{\mu}U^{\dagger})
+function(detU) +\frac{F_\pi}{2} \sum A_a(U_{aa}+U^{\dagger}_{aa}).
\label{3fnlsm}
\end{equation}
The second term will be non-trivial only if there are nine rather
than eight pseudoscalars (i.e. the $\eta^\prime$ is included)
so that $detU \ne 1$.
The quantities $A_a$, proportional to the quark masses, can
be related, by using eq.(\ref{3fnlsm}) or eq.(\ref{lmeson}), to
the meson masses. In this way one can recapture the initially surprising
result~\cite{gor} for the strange to non-strange quark mass ratio: 
\begin{equation}
\frac{A_3}{(A_1+A_2)/2}={\cal O}(25).
\label{symbreak}
\end{equation}
This may be contrasted with the value of about 1.4 expected
in the qualitatively successful non-relativistic quark model. Similarly
one gets~\cite{su3} for the corresponding iso-spin violating quark mass ratio:
\begin{equation}
\frac{A_2-A_1}{(A_1+A_2)/2}={\cal O}(\frac{1}{2}),
\label{isobreak}
\end{equation}
which may be compared with a value about 0.01 expected in the
non-relativistic quark model. These results suggest that the first
three (and especially the first two) quark masses are very small.
Since these quark masses are the source of the intrinsic
chiral $SU(3)_L \times SU(3)_R$ symmetry breaking in the 
Lagrangian, the goodness of the low energy
chiral Lagrangian predictions becomes understandable. In the
 present picture the non-relativistic quark model masses
are identified as arising from the spontaneous breakdown of chiral 
symmetry; they would be non-zero even
 if the true quark masses referred to 
above were zero.
\section{QCD ingredients and Gamma Ray Bursters}

    So far, the arguments leading to eq.(\ref{3fnlsm}) were the ones
used before QCD was discovered; they are general, based only on
flavor symmetries of the strong interaction, and do not need to be
changed. Of course QCD has strengthened their acceptance and led to many
new insights. In the present context, for example, the
violation of the $U(1)_A$ symmetry noted after eq.(\ref{lmeson})
has been clarified. Even though the QCD Lagrangian written at the classical
level obeys this symmetry,
it is illusory since quantum corrections (called the Adler Bell Jackiw 
~\cite{abj} anomaly) spoil the conservation of the would be symmetry current
in the full color gauge theory.

     A related phenomenon may be used to throw some light 
on the ground state of the strong gauge theory and provide the basis for
certain semi-quantitative estimates. Neglecting the light quark masses, the
QCD Lagrangian written at the classical level is invariant under scale
transformations, $x_\mu \rightarrow \lambda x_\mu$. Classsically, this
implies the existence of a corresponding Noether current $D_\mu$
 satisfying the conservation law, $\partial_\mu D_\mu =0$. However
this conservation law is also illusory due to quantum corrections. It
must be replaced by~\cite{scaleanom},
\begin{equation}
\partial_\mu D_\mu =-\frac{\beta(g)}{g}Tr(F_{\mu\nu}F_{\mu\nu}) = H,
\label{H}
\end{equation}
where $\beta(g)$ is the QCD beta function and $F_{\mu\nu}$ is the field
strength tensor of QCD. $H$ is a convenient abbreviation
 for the right hand side. Several authors~\cite{syr,others} have
 considered the possibility of
restricting the effective Lagrangian to mock up this equation.
 This requires the potential term, $V_0$ of eq.(\ref{lmeson})
to contain the scalar glueball field, $H$ and to satisfy,
\begin{equation}
H=dTr(M\frac{\partial V_o}{\partial M}+M^\dagger \frac{\partial V_0}
{\partial M^\dagger}) +4H\frac{\partial V_0}{\partial H})- 4V_0,
\label{euler}
\end{equation}
where $d$ is the scale dimension of $M$. In order to see how this works
let us focus on the simple case in which the quark fields are absent
(pure color gauge theory). We assume that the whole theory is approximated
by the self interactions of the single scalar glueball $H$. Then
the appropriate effective Lagrangian is,
\begin{equation}
{\cal L}_H= -\frac{a}{2}H^{-3/2}(\partial_\mu H)^2-\frac{1}{4}Hln(\frac{H}
{\Lambda^4}),
\label{pureym}
\end{equation}
where $a$ is a dimensionless constant and $\Lambda$ is a QCD energy scale.
The first term is a scale invariant kinetic type term while the second
term provides the solution of the anomaly condition eq.(\ref{euler}).   
It is easily seen that the minimum of the potential in eq.(\ref{pureym})
occurs for $<H>=\Lambda^4/e$, at which point the vacuum energy density
takes the negative value $<V>=-\Lambda^4/(4e)$. In fact, a negative value 
is required in the bag model picture~\cite{bag} of quark confinement.
In that model, if a ``bubble" of non-perturbative, zero energy vacuum is 
made it would tend to collapse
to lower the total energy. However, if quarks
 or gluons are put inside the bubble, their kinetic energy opposes
this effect and results in a stable state. To study the glueball field, 
$h$ in  the simple model of eq.(\ref{pureym}) we set
 $H=\Lambda^4/e +Zh$ and expand. For a given scale $\Lambda$, both $Z$
and $a$ may be gotten by specifying the mass $m_h$. 

    It may be amusing to note that an
 effect of electromagnetism can be 
added to the model through its contribution to the scale anomaly.
 One should replace the second term in eq.(\ref{pureym}) by
$-\frac{1}{4}(H+H_{EM})ln(\frac{H}{\Lambda^4})$, where
$H_{EM}=-\frac{\alpha}{2\pi}{\sum}Q^2_a{\cal A}_{\mu\nu}
{\cal A}_{\mu\nu}$. Here ${\cal A}_{\mu\nu}$ is the electromagnetic field 
strength tensor and the $Q_a$ are the quark charges.
 This would give an estimate of the glueball
decay into two photons in the approximation that it
dominates the energy momentum tensor. Of course this model is  
oversimplified for ordinary QCD since, unlike the case of the 
pseudoscalar mesons, the glueball will not have low mass.

    At this conference many interesting talks were presented
on the treatment of QCD at high density (large
 chemical potential $\mu$). This is a fascinating field~\cite{reviews}
since it gives rise to exotic new phases of matter whose
properties may be computed perturbatively at large $\mu$.
These new phases may be neatly described by effective 
Lagrangians similar to the one describing ordinary QCD.
Possible applications may include matter in compact stars.
As an example, Ouyed and Sannino have recently proposed
a very imaginative model~\cite{os} to understand the 
puzzling behaviors of Gamma Ray Bursters. They argue it is possible
that under the crust of a ``quark star" there is a surface layer
which could be in the so-called 2SC (two flavor superconducting)
phase. In this phase the $SU(3)_C$ color gauge symmetry is
 spontaneously broken to $SU(2)_C$ and a two flavor chiral
symmetry $SU(2)_L \times SU(2)_R$ is preserved in the massless quark 
limit. They go on to show that the lightest physical hadronic states
should actually be glueballs and that the effective Lagrangian
is given by an appropriate modification of eq.(\ref{pureym}).
The gamma ray emissions are born in the $H\rightarrow 2\gamma$
process mentioned above. Finally an ingenious mechanism
in which the state of the surface layer shuttles between the
2SC and QGP (quark gluon plasma) phases is proposed to 
explain the observed episodic character of the gamma ray emissions.

\section{Going away from threshold.}
Gasser and Leutwyler~\cite{gale} have worked out a systematic procedure
for treating eq.(\ref{3fnlsm}) beyond tree level (neglecting the second 
term). Since it is a non-renormalizable theory new counter terms have to 
be added at every order to cancel divergences. There are ten such
 terms at one loop order (Two examples : $c_3Tr(\partial_\mu U 
\partial_\mu U^\dagger
\partial_\nu U\partial_\nu U^\dagger)$ and $c_5Tr(\partial_\mu U
\partial_\mu U^\dagger A(U + U^\dagger))$ where $A=diag(A_1,A_2,A_3)$).
If one counts ${\cal O}(A/F_\pi)={\cal O}(E^2)$, then the tree diagrams
are ${\cal O}(E^2/F_\pi^2)$, the one loop diagrams are ${\cal 
O}(E^4/F_\pi^4)$ and so forth. Thus if $E$ is sufficiently small the 
higher terms in the expansion should be suppressed. This procedure has 
worked very well to correlate a large number of experimental results near 
threshold. It should be remarked that the counter terms also contain
arbitrary finite pieces, which are adjusted to fit the 
data.

    How far up in energy can this chiral perturbation
 approach practically take us? To get an idea consider the experimental 
points shown in Fig. 2 for the real part, $R_0^0(s)$ of the $I=J=0$
pi pi partial wave amplitude. The chiral perturbation series should
essentially give a polynomial fit to this shape,
 which up to 1.0 GeV is crudely 
reminiscent of one cycle of the sine curve. Of course,
\begin{equation}
sin(x)=x-\frac{x^3}{3!}+\frac{x^5}{5!}-\frac{x^7}{7!}+\frac{x^9}{9!}
-\frac{x^{11}}{11!}+\frac{x^{13}}{13!}-\frac{x^{15}}{15!} +\cdots.
\label{sine}
\end{equation}
By using MAPLE or a similar program it will only take a minute to convine 
yourself that the number of terms needed to get a decent approximation to 
one cycle of the sine curve is that just shown. This suggests that 
something like 7 loop chiral perturbation theory would be required
to explain pi pi scattering up to about 1 GeV. An alternative approach,
as the data itself suggests, is to explicitly include resonances.
For example the fit shown in Fig. 2 was computed~\cite{hss} using
a unitarized tree amplitude from a chiral Lagrangian in which scalar 
mesons and vector mesons have been consistently added. The sigma,
which was earlier ``sent to infinity", now turns out to be a broad 
resonance of mass about 560 MeV. There has been a lot of work~\cite{work}
in this area recently and it seems exciting, though still in progress.
Generalization to a full nonet of scalars suggests~\cite{bfss} the light 
scalars actually are more likely to be of $qq{\bar q}{\bar q}$ type~\cite
{j} than of $q{\bar q}$ type. Probably an even better approximation is to 
include~\cite{bfs} mixing between $qq{\bar q}{\bar q}$ and $q{\bar q}$
states.

\begin{figure}[htb]
\begin{center}
\epsfig{file=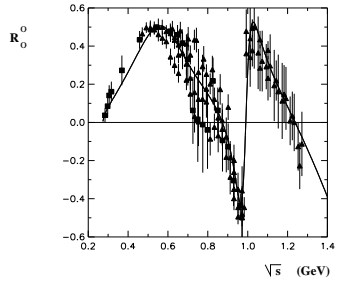,height=1.5in,angle=0}
\caption{Pi Pi scattering amplitude.}
\label{fig:pipi}
\end{center}
\end{figure}

I would like to thank the organizers for a very stimulating  
 conference. This work has been supported in part by
the US DOE under contract DE-FG-02-85ER40231.

\end{document}

%% file: econfmacros.tex
\def\beq{\begin{equation}}
\def\eeq#1{\label{#1}\end{equation}}
\def\eeqn{\end{equation}}

\def\beqa{\begin{eqnarray}}
\def\eeqa#1{\label{#1}\end{eqnarray}}
\def\eeqan{\end{eqnarray}}

\let\bar=\overbar

\def\Dslash{\not{\hbox{\kern-4pt $D$}}}
\def\dslash{\not{\hbox{\kern-2pt $\del$}}}

\def\msb{{\bar{\ssstyle M \kern -1pt S}}}

